\documentclass[aps,pra,onecolumn,groupedaddress,showpacs]{revtex4}

\usepackage{amssymb,amsmath}
\usepackage{graphicx}

\newcommand{\om}{\omega}
\newcommand{\ve}{\varepsilon}

\DeclareMathOperator{\sech}{sech}

\begin{document}


\title{ Nonlinear waves: a review \\
Vector $0\pi$ pulse and the generalized perturbative reduction method}

\author{G. T. Adamashvili}
\affiliation{Technical University of Georgia, Kostava str.77, Tbilisi, 0179, Georgia.\\ email: $adamash77@gmail.com.$ }

\begin{abstract}
In this review, a more general theory of self-induced transparency (SIT) in comparison with the theory of McCall and Hahn is considered. Using the recently developed generalized perturbative reduction method (GPRM) the SIT equations are reduced to vector (coupled) nonlinear Schr\"odinger equations for auxiliary functions.
This approach demonstrate that, unlike McCall and Hahn  SIT theory in which  single-component scalar breather can propagate independently, in the more general theory  of SIT the second derivatives with respect to the spatial coordinate and time of the wave equation play a significant role and describe the interaction of two scalar SIT breathers forming a coupled pair. This is  a vector $0\pi$ pulse of SIT - a two-component  vector breather oscillating with the sum and difference of frequencies and wave numbers. The profile, parameters and properties of this pulse differ significantly from the characteristics of the McCall and Hahn pulses.

Using GPRM it is shown that besides the scalar soliton and the scalar breather, the vector $0\pi$ pulse is also a universal nonlinear wave, arising in virtually all areas of physics where nonlinear phenomena are described by nonlinear equations containing second-order and higher-order derivatives.
A number of such nonlinear equations is presented. Among them almost all well known nonlinear equations, as well as
the general fourth-order nonlinear partial differential equation and the sixth-order generalized Boussinesq-type equations.

\vskip+0.2cm
\emph{Keywords:} Nonlinear waves. Generalized perturbation reduction method. Nonlinear wave equations. Two-component nonlinear solitary waves. Vector breather. Vector $0\pi$ pulse. Solitons. Self-induced transparency.
\end{abstract}


\maketitle

\tableofcontents

\section{Introduction}

Physical phenomena can be conditionally divided into two main kinds. The first kind includes such physical phenomena that can take place only in one (or a few) physical system, requiring specific conditions that can only be realized in that particular system. Examples of such effects are superfluidity and superconductivity, among others. The second kind includes physical phenomena that can take place in many completely different physical systems, regardless of their specific properties and physical nature. Such phenomena are of particular interest because they characterize the general properties of matter. The latter phenomena include the formation of nonlinear solitary waves, which occur in completely different areas of physics for physical quantities with different natures. Examples include the strength of the electric field of an optical wave, the deformation tensor of an acoustical pulse, the deviation of the water surface from the unperturbed state, etc. Nonlinear solitary waves are very interesting physical objects that describe highly excited nonequilibrium states of a system localized in space and time, in which the physical parameters characterizing the system remain unchanged.
 In the theory of nonlinear waves they play as fundamental a role as harmonic oscillations do
in the linear wave theory. The nonlinear waves of an invariable profile are one of the most important demonstrations of
nonlinearity in physical systems.
Unlike linear waves, solitary nonlinear waves can transfer wave energy over considerable distances without significant losses. Nonlinear solitary waves have long been a subject of study, and a significant number of works have been published; nevertheless, they are still relevant and their intensive research continues. These waves were first discovered on the surface of liquids, but were subsequently also investigated in other areas of physics, including optics, condensed matter, plasma, field theory, acoustics, metamaterials, quantum dots, graphene, etc. These nonlinear waves in different media and different physical conditions have identical properties, although they are solutions of completely different nonlinear partial differential equations, such as the Korteweg de Vries equation, the Boussinesq equation, the Benjamin-Bona-Mahony equation, the Bloch-Maxwell system of equations, the Hirota equation, the nonlinear Schr\"odinger equation (NSE), the sine-Gordon equation (SGE) and many other equations (see, for instance, Refs.[1-8]).

Among nonlinear solitary waves, single-component (scalar) and two-component (vector) waves are the most common [9, 10]. They differ significantly from each other in terms of their shapes, parameters and physical properties. Single-component waves in the process of propagation retain not only their energy but also their profile. Single-component nonlinear solitary waves include a scalar soliton, a scalar breather, and other scalar nonlinear waves. Two-component solitary nonlinear waves are more complex formations and have an internal structure. Two-component waves include a vector breather, or a breather molecule [11-13]. A vector breather is a bound state of two scalar breathers that oscillate at different frequencies, have the same velocities, propagation directions, and polarizations, or have mutually perpendicular polarizations, for example, for a waveguide mode. In the process of propagation, the scalar breathers that make up the vector breather interact with each other and exchange energy, although they propagate as a whole.

For the analytical description of single-component and two-component nonlinear solitary waves, the use of various mathematical methods is required. This is because more functions and parameters are required to study two-component nonlinear waves than to study single-component nonlinear waves. For example, the perturbative reduction method (PRM)[14, 15] is often used to study single-component nonlinear waves, utilizing one auxiliary complex function and two real parameters. Via this method, it is impossible to study two-component nonlinear waves. For two-component nonlinear waves, a generalized version of the PRM  was developed [16-18]; this uses two complex auxiliary functions and eight parameters.

Single-component and two-component nonlinear solitary waves can be formed in both different and identical physical systems. For the implementation of each, appropriate physical conditions are required. Some effects cannot be described using single-component nonlinear waves, and the concept of two-component waves becomes necessary. The most striking example of such effects is self-induced transparency (SIT)[19, 20]. After the discovery of this effect, for half a century it was believed that the main SIT pulses are a scalar soliton ($2\pi$ pulse) and a scalar breather (scalar $0\pi$ pulse).  They are called McCall and Hahn  pulses.

However, after the development of a generalized perturbative reduction method (GPRM)[16-18], enabling the study of two-component waves, it was discovered that the physical picture of the SIT is qualitatively different. In the theoretical description of SIT, the second derivative of the strength of the electric field of the pulse concerning the spatial coordinate and time in Maxwell's wave equation plays a fundamental role. However, in the traditional theory of SIT formulated by McCall and Hahn and later by many others, the second derivatives were neglected, or were taken into account according to the perturbation theory only as small corrections of the first derivatives [21-24].

 Through the GPRM it became possible to consistently take into account the role of second derivatives in Maxwell's wave equation. It was shown that the second derivatives describe the interaction between scalar breathers and the formation of their bound state. As a result, it was found that the SIT scalar breathers pair form a vector breather and, therefore, the main SIT pulses are, along with the scalar soliton ($2\pi$ pulse), a vector $0\pi$ pulse.

 The vector $0\pi$ pulse is a vector breather that consists of a bound state of two scalar single-component breathers of the same polarization, one of which oscillates with the sum, and the second with the difference, of frequencies and wave numbers [16-18, 25]. In this review, by vector breather we will mean exactly such vector $0\pi$-pulse.

 Subsequently, such a vector breather was studied in acoustic SIT, and then in many other areas of physics for quantities of a completely different nature, which are described by such nonlinear equations as, for instance,  the Korteweg de Vries equation, the Boussinesq equation, the Benjamin-Bona-Mahony equation, etc. To study vector breather in these equations, the GPRM was also used
[26-31].

For the analysis of the nonlinear solitary waves the pulse width $T$ plays an important role. We can consider two types of pulses. The first are very specific the few-cycle ultra-short duration pulses  and the second are relatively wider pulses, for which the inequality $\omega T >> 1$ is valid, where $\omega$ is the carrier wave frequency. Slowly varying envelope approximation (SVEA) is usually used for the last type of pulse [19, 32-34]. For such kind pulses, the function $u(z, t)$  can be represented as:
\begin{equation}\label{UE}
u(z,t)=\sum_{l=\pm 1}\hat{u}_{l}(z,t) Z_l,
\end{equation}
where $u(z,t)$  is a real function of spatial coordinate $z$ and time $t$ and represents the nonlinear pulse profile, $Z_{l}= e^{il(kz -\om t)}$ is the fast oscillating function, $\hat{u}_{l}$  is the slowly varying complex envelope function, which satisfies inequalities
\begin{equation}\label{swa}
 \left|\partial_{t} \hat{u}_{l}\right|\ll\omega
|\hat{u}_{l}|,\;\;\;\left|\partial_{z} \hat{u}_{l}\right|\ll k|\hat{u}_{l}|,
\end{equation}
$k$ is the wave number of the carrier wave. For the reality of $u$, we suppose that: $ \hat{u}_{+1}= \hat{u}^{*}_{-1}$.

The physical quantity $u(z,t)$ can have different natures. Examples include the electric field strength of an optical wave, the deformation tensor of an acoustic pulse, the deviation of the water surface from the undisturbed state, the area of pulse envelope, etc.

The purpose of this review is to demonstrate, using the recently developed GPRM (ansatz), that more general SIT theory differs significantly from the original McCall and Hahn SIT theory. Specifically, to prove the significant role of second-order and higher-order derivatives in wave equations in describing nonlinear wave processes. To demonstrate that the fundamental pulse of SIT is a vector $0\pi$ pulse, but not the scalar $0\pi$ pulse of McCall and Hahn.
To show that, along with the scalar soliton and the scalar breather, the vector $0\pi$ pulse is also a universal nonlinear wave, encountered in virtually all areas of physics and is a solution to a number of nonlinear equations for completely different physical quantities.

The remainder of this review is organized as follows: Section II is devoted to the SIT theory of McCall and Hahn. In Section III, using the GPRM, we consider SIT theory more generally. To study the role of second derivatives with respect to the spatial coordinate and time in the Maxwell wave equation and transform this equation into vector (coupled) nonlinear Schr\"odinger equations (VNLEs) for auxiliary functions. This will yield a two-component vector $0\pi$ pulse of the SIT. Section IV presents solutions of two-component vector breathers oscillating with the sum and difference of frequencies and wave numbers (SDFW) for a number of various nonlinear equations. In Section V, we discuss the obtained results. Section VI presents the abbreviations used. Finally, Section VII and the corresponding appendices contain information on scalar solitons and scalar breathers, as well as detailed derivations and solutions of the equations discussed in the text.

\section{Self-induced transparency. McCall and Hahn pulses}

\label{MHS}

The optical resonant nonlinear waves of the stable profile can be created by means of the McCall-Hahn mechanism, i.e., from the nonlinear coherent
interaction of an optical pulse with ensembles of resonant optical  atoms in a medium, such as gaseous atoms, impurity atoms in
solids or semiconductor quantum dots (SQDs), when the conditions $\omega T>>1$ and $T<<T_{1,2}$ are satisfied. Here $T_{1}$ and $T_{2}$ are the longitudinal and transverse relaxation times of the resonant atoms or SQDs. Exploring the evolution of a coherent light pulse in a resonant absorbing medium, McCall and Hahn observed the amazing phenomenon of anomalously low energy loss when the pulse power exceeds some critical value. This effect was called SIT [19]. The intensity of the pulse interaction with atomic systems or SQDs is characterized by the area of the pulse envelope. On the other hand, the area of the pulse envelope determines the type of nonlinear wave. Following the McCall-Hahn area theorem when the area of the pulse envelope $\Psi$ exceed $\pi$, $2\pi$ hyperbolic secant pulse is formed and for low intensity pulses, if $\Psi<<1$, then $0\pi$ pulse is generated. The $4\pi$, $6\pi$, ... pulses in the process of propagation are divided into a discrete sequence of $2\pi$ pulses. Consequently, in resonantly absorbing media McCall and Hahn's $2\pi$  and  $0\pi$ pulses are the basic pulses of SIT. These are completely different single-component scalar nonlinear waves with different properties and conditions of existence, which have been intensively studied under different physical conditions and in various materials
[32-43] (see also Appendix A).

The physical explanation of the SIT effect is based on the representation of the absorber by an ensemble of two-level atoms whose evolution is caused by induced processes due to the interaction with the coherent light pulse. In general the theory of the interaction of electromagnetic radiation with an ensemble of two-level atoms is based on the Bloch equations for atoms and the Maxwell equations for the classical electromagnetic field (see Appendix B).

There are several well-known methods which can be used to solve undamped SIT equations (B5) and (B6) and to investigate the scalar single-component $2 \pi n$ pulses of SIT in attenuator media ($n=0,1,2...$). For instance, the factorization of the polarization and to introduce the dipole "spectral response" function [19, 32-34, 43], multiple scale method [4, 44], and the PRM [14].

From the undamped-Bloch-Maxwell equations  (B5) and (B6), many of the principal theoretical results of SIT were originally obtained by McCall and Hahn, using quite simple methods, in particular, the fact that certain solutions of this system of equations, known as $2\pi$ pulses, did indeed have solitonlike properties [19, 45].

The  effect of SIT can be described using the undamped Bloch-Maxell equation   (B5) and (B6) in the presence of phase modulation and by SGE (13) in the absence of phase modulation. Despite the fact that the complete $n$-soliton solutions of these equations, by means of the inverse scattering transform (IST), are obtained [3-5, 44-47].

The results obtained by IST are nevertheless very limited. This is due to the fact that a number of very important limitations were allowed when deriving the equations (B5), (B6) and (13). In particular, the second derivatives with respect to spatial coordinate and time, terms describing Markovian and non-Markovian transverse and longitudinal relaxation processes, terms taking into account the interaction of wave with conduction electrons, terms taking into account pumping, the transverse structure of the wave, etc. were neglected. For an adequate description of the SIT, it is necessary to take these effects into account, especially since some of them significantly change the physical picture of the SIT, and in particular, taking into account the second derivatives in Maxwell's equations can lead to qualitatively new results (see Section 3).
Special interest express small intensity  nonlinear waves which met in a lot of different physical phenomena. To take into account the role of second derivatives in SIT equations, it is necessary to use other mathematical approach, namely GPRM.

\section{The generalized perturbative reduction method.
Vector $0\pi$ pulse. \;\;\;\;\;\;\;\;\;\;\;\; \;\;\;\;\;\;\;\;\;\;\;\;\;\;\;\;\;\;\;\;\;\;\;\;\;\;\;\;\;\;\;\;\;A more general theory of self-induced transparency}

\label{GPRMS}
In theory of nonlinear waves and, in particularly, in the theory of SIT, usually in the wave equation (B4) for slowly envelope amplitudes $\hat{E}_{l}$, it is sufficient to take into account only the first derivatives terms of $\hat{E}_{l}$ with respect to the spatial coordinates $\frac{\partial \hat{E}_{l}}{\partial z}$ and time  $\frac{\partial \hat{E}_{l}}{\partial t}$. The corresponding second derivatives terms $\frac{\partial^{2} \hat{E}_{l}}{\partial z^{2}}$ and $\frac{\partial^{2} \hat{E}_{l}}{\partial t^{2}}$   in the nonlinear wave equation (B4) usually have been neglected. In the frame of such approximation have been considered soliton and breather  solutions of the wave equation in a lot of physical situations and various materials. Such approach have been widely used starting from the first study of the theory of SIT [19, 32-36, 43-46].

But when we neglect the second derivative terms in the nonlinear wave equation (B4), at the same time arise the question: what kind of effects we ignore when we neglect the second derivative terms in Eq.(B4) and can we obtain the qualitatively new physical results when we  take into account these terms. In order to answer of this question  is necessary besides of the first derivative terms to keep also the second derivative terms in Eq.(B4) and analyze this equation in more general case.

Influence of  the second-order derivatives  in the spatial coordinate  and time  of the envelope of the strength of electric field  of SIT pulse can be considered by means of using various expansion methods, in particular, the PRM [4,14, 44], and  the contribution of  the second-order derivatives leads only to small corrections on the parameters of the single-component nonlinear waves (see, for instance, Refs.[21-24]).

However, it was recently shown that considering the second-order derivatives in the spatial coordinate  and time in the wave equation (B4), we  can also obtain qualitatively new results.
It becomes possible if we consider the  SIT phenomenon and the corresponding equations  in a more general case, by applying the GPRM developed in Refs.[16-18].

This method makes it possible to transform some nonlinear partial differential equations and in particular, the Eq.(B4) to VNSEs for two complex auxiliary functions and it provides the solution to these equations in the form of the two-component vector $0\pi$ pulse - vector breather oscillating with the SDFW. Using this approach (anzatz), the complex envelope function $\hat{u}_{l},$  Eq.(1) can be represented as [16-18, 48]:
\begin{equation}\label{gprm}
\hat{u}_{l}(z,t) =\sum_{\alpha=1}^{\infty}\sum_{n=-\infty}^{+\infty}\varepsilon^\alpha
Y_{l,n} f_{l,n}^ {(\alpha)}(\zeta_{l,n},\tau),
\end{equation}
where $\varepsilon$ is a small parameter,
$$
Y_{l,n}=e^{in(Q_{l,n}z-\Omega_{l,n}
t)},\;\;\;\zeta_{l,n}=\varepsilon Q_{l,n}(z- {{v}_{g;}}_{l,n} t),\;\;\;\tau=\varepsilon^2 t,\;\;\;
{{v}_{g;}}_{l,n}=\frac{d\Omega_{l,n}}{dQ_{l,n}}.
$$

Such a representation allows us to separate from $\hat{u}_{l}$ the still more slowly changing
quantities $f_{l,n}^{(\alpha )}$. Consequently, it is assumed that the quantities $\Omega_{l,n}$, $Q_{l,n}$ and $f_{l,n}^{(\alpha)}$ satisfies the inequalities for any $l$ and $n$:

\begin{equation}\label{gprm1}
\omega\gg \Omega_{l,n},\;\;k\gg Q_{l,n},\;\;\;
\left|\partial_{t} f_{l,n}^{(\alpha )}\right|\ll \Omega_{l,n} \left|f_{l,n}^{(\alpha)}\right|,\;\;\left|\partial_{z}
f_{l,n}^{(\alpha )}\right|\ll Q_{l,n} \left|f_{l,n}^{(\alpha )}\right|.
\end{equation}

Unlike the usual, standard form of the PRM [14], in the perturbation expansions Eqs.(3) and (4), the quantities $\Omega_{l,n}$, $Q_{l,n}$, ${{v}_{g;}}_{l,n}$, $\zeta_{l,n}$ and $f_{l,n}^{(\alpha)}$  are depends from the indexes $l$ and $n$.
In the special case when these quantities are not depends from the indexes $l$ and $n$ the expansionsEqs.(3) and (4) are transformed to the well known standard perturbation reduction expansion [14].

Substituting Eq.(3) for the complex envelope function ${\Psi}_{l}(z,t)$ Eq.(C1) into Eq.(B4)  we determine the connection between the oscillating parameters $\Omega_{l,n}$  and $Q_{l,n}$  (see Appendix C)
\begin{equation}\label{2diss}
l n (2 \omega \Omega_{l,n} -2 k v^{2}  Q_{l,n} -  \frac{\beta^{2}}{\Omega_{l,n}}) = v^{2} Q^{2}_{l,n} -  \Omega^{2}_{l,n}
\end{equation}
and the VNSEs for the auxiliary function $U_{\pm}=\ve f_{+1,\pm 1}^ {(1)}$ in the following form
\begin{equation}\label{cnse2}
 i  (\partial_{t} U_{\pm}+ v_{\pm} \partial_{z} U_{\pm})
+  p_{\pm} \partial_{zz} U_{\pm} + {q}_{\pm} |U_{\pm}|^{2}U_{\pm} +r_{\pm} |U_{\mp}|^{2} U_{\pm}=0,
\end{equation}
where
\begin{equation}\label{koef}
v_{\pm}={{v}_{g;}}_{+1,\pm 1},\;\;\;\;\;\;\;\;\;\;\;
p_{\pm}=-\frac{{H}_{+1,\pm 1}}{{h}_{+1,\pm 1}Q^{2}_{\pm }},\;\;\;\;\;\;\;\;\;\;\;\;\;\;\;\;
{q}_{\pm}=-\frac{ \beta^{2}}{2 {h}_{+1,\pm 1}},\;\;\;\;\;\;\;\;\;\;\;\;\;\;\;\;
r_{\pm}=q_{\pm}(1  - \frac{\Omega_{\mp } }{\Omega_{\pm }}),
$$$$
\Omega_{\pm }=\Omega_{+1,\pm 1}=\Omega_{-1,\mp 1},\;\;\;\;\;\;\;\;\;\;\;\;\;\;\;\;
Q_{\pm }=Q_{+1,\pm 1}=Q_{-1,\mp 1}.
\end{equation}

The solutions of the Eq.(5) depend  on the variables $l=\pm 1$ and $n=\pm 1$. Therefore we have four equations for the quantities  $\Omega_{+1,+1}$,  $\Omega_{+1,-1}$,  $\Omega_{-1,+1}$ and  $\Omega_{-1,-1}$ for corresponding values of the quantities $Q_{l,n}$. But since  Eq.(5) depends only on the products of variables $l\;n$,\, it reduces to two independent equations:
\begin{equation}\label{2diss2}
\pm(2 \omega \Omega_{\pm} -2 k v^{2}  Q_{\pm} -  \frac{\beta^{2}}{\Omega_{\pm}}) = v^{2} Q^{2}_{\pm} -  \Omega^{2}_{\pm}.
\end{equation}

The terms on the right hand side of the  Eq.(5)   $v^{2} Q^{2}_{l,n}$ and $ \Omega^{2}_{l,n}$, or $v^{2} Q^{2}_{\pm}$ and $ \Omega^{2}_{\pm} $ in Eq.(8), arise in these equations from the second derivative terms of the wave equation (B4) $v^{2} \frac{\partial^{2} \hat{E}_{l}}{\partial z^{2}}$ and $\frac{\partial^{2} \hat{E}_{l}}{\partial t^{2}}$. If we neglect these second derivative terms in Eq.(B4), as usually have made in the theory of SIT [19, 32-36, 43-46], the right hand sides of the  Eqs.(5) and (8) will be equal to zero and the parameters $Q_{l,n}$ and $\Omega_{l,n}$ ($Q_{\pm}$ and $\Omega_{\pm}$)  will not depends from the indexes $l$ and $n$.
Consequently, under this condition from the Eqs.(7) and (8) we obtain that $\Omega_{+}=\Omega_{-},\;$ $\;r_{\pm}=0$ and the VNSEs Eq.(6) will be disconnected to the two independent scalar NSEs (10). As result, we obtain two independent single-component breathers which will be propagating separately to each other.

From this facts we can make the very important conclusion that the first-order derivatives in the spatial coordinate  and time in the wave equation (B4)  describes of the formation of the single-component solitons and the single-component breathers and the second-order derivatives in the spatial coordinate  and time in the wave equation(B4)
describes of the interaction between two single-component breathers and provide the formation of a bound state of these breathers, i.e. the two-component vector breather (vector $0\pi$ pulse). This situation is somewhat similar to the formation of a Cooper pair of electrons, which forms at low temperatures in a superconductor, providing superconductivity.

The VNSEs (6) have been analyzed widely for several years (for instance [9, 50] and references therein). In the simplest case when all coefficients are equal to each other, the system of two equations (6) is solved exactly using IST [51] (see also Appendix C).

Substituting the solutions of Eq.(6) into Eq.(1) and using the notation $u(z,t)=E(z,t),$ we obtain for the $x$ component of the electric field strength $E$ of the pulse the two-component vector $0\pi$ pulse - vector breather oscillating with the SDFW of the SIT in the following form:
\begin{equation}\label{VB1}
E(z,t)= \frac{\hbar }{{\mu}_{12}\; b\; T}\sech(\frac{t-\frac{z}{V_0}}{T})\{ w_{+} K_{+} \sin[(k+\kappa_{+})z -(\om + w_{+}) t]
- w_{-} K_{-} \sin[(k-\kappa_{-})z -(\om -w_{-})t]\},
\end{equation}
where   $w_{\pm}=\Omega_{\pm}\pm \omega_{\pm},$\;\;\;$\kappa_{\pm}=Q_{\pm}\pm k_{\pm}.$

Eq.(9) describes the nonlinear two-component wave that consists of the two breathers.
One of the breather oscillates with the sum  of the frequencies $\om +w_{+}$ and wave numbers $k+\kappa_{+} $ and the second breather with the difference of the frequencies $\om -w_{-}$ and wave numbers $k-\kappa_{-} $.  The parameters of the nonlinear wave by the equations (7), (8), (C3) and (C6) are determined. The corresponding plot of the two-component vector breather oscillating with the SDFW at a fixed value of the $z$ coordinate
is presented in Fig.1.

\begin{figure}[htbp]
\includegraphics[width=0.44\textwidth]{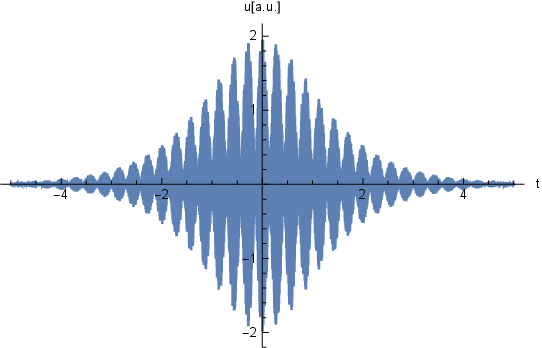}
\caption{ Plot of the vector $0\pi$ pulse - a two-component  vector breather oscillating with the SDFW
 at a fixed value of the z coordinate. The dark areas of the figure indicate
the oscillation regions.}
\label{fig1}
\end{figure}

It should be noted hat the analyze of Eq.(B4) by means of expansion (3) is valid only in case when  $\Psi_{l}$ is complex function, i.e. when the phase modulation of the nonlinear pulse take place. Otherwise, when the phase modulation is absent, the functions $\Psi_{l}$  and $\hat{E}_{l}$ are transformed to the real functions  $\Psi$  and $\hat{E}$ which are independent of the index  $l$. Under these conditions the wave equation (B4) impossible to transform to the VNSEs
(exception for real function, see Ref.[16]).

After investigating the influence of second-order derivatives in the wave equations using GPRM,  an important conclusion can be made: the basic pulse of SIT along with the $2\pi$ pulse is also the vector $0\pi$ pulse, but not the scalar $0\pi$ pulse. The scalar   $0\pi$ pulse of SIT is an approximation which can be considered when we ignore the second derivatives in SIT equations,  when the phase modulation is absent, or use methods (for instance, PRM [4, 14, 44] which do not allow the consideration of two-component waves.

In soliton theory terminology, the single-component $2\pi$-pulse and single-component $0\pi$-pulses of the SIT are a scalar soliton and a scalar breather, and the two-component vector $0\pi$-pulse is a vector breather.

 The same phenomenon, namely the formation of a vector breather consisting of two breathers oscillating with the SDFW, can be observed in completely different areas of physics, for quantities of completely different physical nature, which are described by different nonlinear equations. A number of such nonlinear equations will be presented in  Section 4.

\vskip+0.7cm

\section{Vector $0\pi$ pulse and  various nonlinear equations}

There exist nonlinear partial differential equations for a real function $u(z,t)$ and a complex function $V(z,t)$, where $u=V+V^{*}$.

\vskip+0.2cm

\textbf{Nonlinear equations for complex function $V$}

\subsection{The scalar nonlinear Schr\"odinger equation [52]}

The scalar NSE [3-6, 44]
 \begin{equation}\label{nse}
 i V_{ t}+ \rho V_{zz} +\delta |V|^{2} V  =0
\end{equation}
is one of the basic equation for studying the solitary waves of stable profile of any nature in different physical fields of research: in optics, acoustics, magnetics, fluid dynamics, quantum electronics, particle physics, plasma physics and etc.

\subsection{The complex modified Korteweg-de Vries equation and the Hirota equation [53, 54]}

The Hirota equation has the following form [53-56]
\begin{equation}\label{hir}
 i V_{ t}+ \rho V_{zz} +\delta |V|^{2} V  +i\sigma V_{zzz}    +i 3 \alpha |V|^{2} V_{z} =0,
\end{equation}
where $\alpha,\;\rho,\;\sigma,\; \delta$ are the real constants and for which the condition
\begin{equation}\label{coefq}
\alpha \rho=\sigma \delta
\end{equation}
is valid.

When  $\alpha=\sigma=0$,  the Hirota equation (11) can be reduced to the scalar NSE Eq.(10).
But,  when $\rho=\delta=0$,  Eq. (11) is transformed to the complex modified Korteweg-de Vries equation
\begin{equation}\label{kdv}\nonumber
 V_{ t} +\sigma V_{zzz} + 3 \alpha |V|^{2} V_{z}  =0.
\end{equation}

Sometimes, it is convenient  rewritten the Hirota equation (11)  to the following form [57]
\begin{equation}\label{Hir3}\nonumber
 i V_{ t} -\alpha_{2}( V_{zz} +2 |V|^{2} V)  +i\alpha_{3}( V_{zzz}    +6|V|^{2} V_{z}) =0,
\end{equation}
where
$$
\alpha=2\alpha_{3},\;\;\;\;\;\;\;\;\;\;\;\;\;\delta=-2\alpha_{2},\;\;\;\;\;\;\;\;\;\;\;\;\;\rho=-\alpha_{2},\;\;\;\;\;\;\;\;\;\;\;\;\;\sigma=\alpha_{3},
$$
in such a way that the constraint Eq.(12) is satisfied.

\vskip+0.8cm

\textbf{Nonlinear equations for real function $u$}

\subsection{Sine-Gordon equation and Klein-Gordon equation [16, 58]}

A nonlinear coherent interaction of an optical pulse with resonant optical atoms, is governed by the Bloch-Maxwell equations. When the Rabi frequency of the wave is real and  the longitudinal and transverse relaxations of the optical atoms are ignored, these equations are reduced to the SGE [3-5, 44]
\begin{equation}\label{sge}
 u_{ tt}-v u_{ zz} =-\alpha_{0}^{2} \sin u,
\end{equation}
 where  $v$ and $\alpha_{0}^{2}=\beta^{2}/2\omega$  are the real constants (see Appendix B).

 The SGE describes the geometry of surfaces with Gaussian curvature, the Josephson transition, dislocations in crystals, waves in ferromagnets associated with the rotation of magnetization, the properties of elementary particles and the nonlinear phenomena in the various fields of physics: optics, acoustics, plasma physics, semiconductor quantum dots, graphene, optical and acoustical metamaterials, and others [2-5].

Among solitary waves, the nonlinear waves  a relatively low amplitude
$u<<1 $
are considered quite often. Under this condition SGE (13) is reduced to the nonlinear Klein-Gordon equation [2]
\begin{equation}\label{kg}
u_{ tt}-v u_{ zz} =-\alpha_{0}^{2}u +\frac{\alpha_{0}^{2}}{6} u^{3} -\mathcal{O}( u^5).
\end{equation}

\subsection{Born-Infeld equation [63]}

There is a nonlinear modification of the Maxwell wave equation, which includes the so-called Born-Infeld nonlinearity. As a result, the Born-Infeld equation is obtained, which describes the properties of particles, the interaction of electromagnetic waves in nonlinear electrodynamics, various phenomena in the field theory, the theory of strings, some of the atomic experiments and has the form [2, 59-64]
\begin{equation}\label{bi}
u_{ tt}-C u_{ zz}=-  A (u_{t})^{2} u_{ zz}  - \sigma (u_{z})^{2} u_{ tt} + B u_{t} u_{z} u_{zt},
\end{equation}
where
$A,\;B,\;C$ and $\sigma $ are the real constants.
Eq.(15) has solutions in the form of the solitary waves $u=\Phi(x+t)$ and $u=\Phi(x-t)$, where  $\Phi$ is arbitrary function.

\subsection{Born-infeld equation and Sine-Gordon equation [65]}

Although the Born-Infeld equation and SGE describes a fairly wide range of absolutely different phenomena in various fields of physics, nevertheless, both of these equations are nonlinear Maxwell wave equations but with different nonlinear terms. Consequently arises the natural question: is it possible to describe all the effects connected with these two equations in a unified way, by means of one more general equation?

To answer this question, the following more general nonlinear wave equation is considered [65]
\begin{equation}\label{bisg}
u_{ tt}-C u_{ zz}=-\alpha_{0}^{2} \sin u  -  A (u_{t})^{2} u_{ zz}  - \sigma (u_{z})^{2} u_{ tt} + B u_{t} u_{z} u_{zt},
\end{equation}
where $\alpha_{0}^{2},\;$ $A,\;B,\;$ $ C,$ and $\sigma $ are the real constants.

Eq.(16) in particular cases goes into the SGE and the Born-Infeld equation.
Indeed, when the condition  $A=B=\sigma=0 $  is fulfilled, Eq.(16) is transformed into the SGE (13).
In case when the condition $\alpha_{0}^{2}=0$ is fulfilled, Eq.(16) is reduced to the Born-Infeld equation (15).

Under the condition $u<<1$ the last nonlinear wave equation is reduced to the form
\begin{equation}\label{bikg}
u_{ tt}-C u_{ zz}=-  A (u_{t})^{2} u_{ zz}  - \sigma (u_{z})^{2} u_{ tt} + B u_{t} u_{z} u_{zt}-\alpha_{0}^{2} u +\frac{\alpha_{0}^{2}}{6} u^{3} -\mathcal{O}( u^5).
\end{equation}

\subsection{Boussinesq equations[66]}

\subsubsection{Boussinesq equation. Version 1 }

The nonlinear properties of shallow water waves can be modeled by various nonlinear partial differential equations. These include
the Benjamin-Bona-Mahony equation, the Korteweg-de Vries equation and different versions of the Boussinesq equations, among others []7, 26, 67-71].

The generalized Boussinesq equation has the form [70, 72]
\begin{equation}\label{eq1}\nonumber
 u_{ tt}-C u_{ zz} - D   u_{ zzzz} + G  (u^{n})_{zz} =0,
\end{equation}
where $C,\;D $ and $G $ are arbitrary constants.

When $n = 3$, the last equation reduces to the cubic Boussinesq equation
\begin{equation}\label{e2}
 u_{ tt}-C u_{ zz} - D   u_{ zzzz} + G  (u^{3})_{zz} =0.
 \end{equation}

Eq.(18) is a well-known model of dispersive nonlinear waves which describes properties of nonlinear solitary waves in a one dimensional lattice and in shallow water under gravity, in geotechnical engineering practice. The Boussinesq equation is also used in the analysis of many other phenomena [73, 74].
There are several well-known methods which can be used to solve the cubic Boussinesq equation. The tanh method, the variational iteration method, and several other
approaches to solve Eq.(18) and to analyze the solitary waves have been applied [72, 75, 76].

\subsubsection{Boussinesq equation. Version 2}

Sometimes, another form of the cubic Boussinesq equation when the term $u_{zzzz}$ is replaced by the term $u_{zzzt}$  is also considered [29, 77] (see, Eq.(23)

\begin{equation}\label{eq3}
 u_{ tt}-C u_{ zz} - D   u_{ zzzt} + G  (u^{3})_{zz} =0.
\end{equation}

\subsection{Boussinesq-type equation [29, 77]}

The modified improved Boussinesq equation

\begin{equation}\label{bim}
 u_{ tt}-C u_{ zz} -   u_{ zztt} + a  (u^{3})_{zz} =0.
\end{equation}
is one more modified type of Boussinesq equations for the modelling water-wave problems in a weakly dispersive medium such as surface waves in shallow waters or ion acoustic waves [78].

The generalized modified Boussinesq equation is [79]
\begin{equation}\label{gmbe}
 u_{ tt} - \vartheta  u_{ tt  \xi \xi}=  [f(u)]_{\xi \xi} ,
\end{equation}
where $\vartheta$ is a constant, $f(u)$ is an arbitrary nonlinear function, and $\xi$ is the the spatial coordinate. Eq.(21) describes the wave propagation of elastic rods and also the nonlinear lattice modes, iron sound waves, and vibrations in a nonlinear string.

In the system of coordinates moving along  axis $\xi$ with velocity $V$, Eq.(21) is transformed into the following equation:
\begin{equation}\label{MBA1}
  u_{ tt}+V^{2}u_{ zz}-2 V u_{ tz} +V^{2 } \vartheta u_{ zzzz} - 2 V \vartheta u_{zzzt}+\vartheta u_{zz tt}= [f(u)]_{zz},
\end{equation}
where
$$
\xi=z-V t,\;\;\;t=t.
$$

We consider the following nonlinear cubic Boussinesq-type  equation which can incorporate the above presented different versions of the Boussinesq-type equations (18), (19), (20) and (22)
\begin{equation}\label{gb}
 \alpha u_{ tt} +\beta u_{ zz}+ \gamma u_{ zt} +\delta u_{ zzzz} +\mu u_{ zzzt}+\nu u_{zz tt}=- G   (u^3)_{zz},
\end{equation}
where we suppose that the nonlinear arbitrary function
\begin{equation}\label{fg}
f(u)=-G u^{3},
\end{equation}
where $G$ is arbitrary constant.

The cubic Boussinesq-type equation (23) is a general equation which united several special cases of the cubic Boussinesq equations.

Indeed, when  the coefficients in Eq.(23) satisfied the conditions
\begin{equation}\label{c1}\nonumber
\alpha=1,\;\;\beta=-C,\;\; \delta=-D,\;\;\gamma=\mu =\nu =0,
\end{equation}
this equation was transformed to Eq. (18).

In case when the coefficients in Eq.(23) satisfied the conditions
\begin{equation}\label{c2}\nonumber
\alpha=1,\;\;\beta=-C,\;\; \mu=-D,\;\;\gamma=\delta =\nu =0,
\end{equation}
than Eq.(23) coincided with  Eq.(19).

Under the condition when
\begin{equation}\label{c3}\nonumber
\alpha=1,\;\;\beta=-1,\;\;\gamma=\delta=\mu=0,\;\;\nu=-1,\;\; a=-G,
\end{equation}
than Eq. (23) is reduced to Eq.(20).

If the coefficients in Eq.(23) satisfied  the conditions
\begin{equation}\label{c4}\nonumber
 \alpha=1,\;\;\beta=V^{2},\;\;\gamma=-2V,\;\;\delta=V^{2} \vartheta,\;\;\mu=-2 V \vartheta,\;\;\nu=\vartheta,\;\;f=-G u^{3},
 \end{equation}
than Eqs.(22) and (23) coincided.

The Boussinesq-type equations (18) - (23) can describe nonlinear waves in various fields of physics and the wave of different nature. Consequently, the meaning of the variables included in these equations is also different. For instance, in hydrodynamics, the function $u$ describes the deviation of the water surface from the unperturbed state, and the coefficients are expressed by means of the parameters of the free fall acceleration and the depth of the unperturbed fluid [2, 8].

\subsection{The sixth-order generalised
Boussinesq-type equations [80]}

\subsubsection{The sixth-order generalised
Boussinesq-type equation. Version 1}

All nonlinear partial differential equations investigated so far, in which two-component vector $0\pi$ pulses were considered, contain partial derivatives concerning spatial coordinates and time, including the first to the fourth order. However, there are equations describing important physical phenomena containing partial derivatives with respect to spatial coordinates and time of a higher order. These include, in particular, nonlinear partial differential equations containing partial derivatives concerning spatial coordinates and time of the sixth order. It is noteworthy to express the sixth-order generalised Boussinesq-type equations among them. These equations are used to study the propagation of the nonlinear solitary water waves with surface tension [81-86].

The two versions of the sixth-order generalised Boussinesq-type equations are considered.

The first version of the sixth-order generalised Boussinesq-type nonlinear wave equation has the following form:

\begin{equation}\label{b61}
\alpha u_{tt} + \beta u_{zz} + \delta u_{zzzz} + \nu u_{zztt}+ S u_{zzzztt} =[f(u)]_{z,z}
\end{equation}
where  $\alpha$, $\beta$, $\delta$, $\nu$, and $S$ are arbitrary constants. $f(u) $ is an arbitrary nonlinear function.

\subsubsection{The sixth-order generalised
Boussinesq-type equation. Version 2}

The second version of the sixth-order generalised Boussinesq-type nonlinear wave equation has the form:
\begin{equation}\label{b62}
\tilde{\alpha} u_{tt}  + \tilde{\beta} u_{zz} +\tilde{ \delta }u_{zzzz}  + \tilde{\nu} u_{zztt}+ \tilde{S} u_{zzzzzz}= [f(u)]_{z,z}
\end{equation}
 where $\tilde{\alpha}$, $\tilde{\beta}$, $\tilde{\delta}$, $\tilde{\nu}$, and $\tilde{S}$ are arbitrary constants.

Eqs.(25) and (26) differ from each other in terms of the sixth-order partial derivatives concerning spatial coordinates and time. In equation (25), the sixth-order derivative term has the form $u_{zzzztt}$, while in equation (26), it has a different form  $u_{zzzzzz}$.

The nonlinear term $f(u)$ in the sixth-order generalised Boussinesq-type equations (25) and (26) very often has a nonlinearity of the second-order or third-order form [81-86].
We consider situations when the nonlinear arbitrary function has the form: Eq.(24)[80].

\subsection{The nonlinear dispersive Benjamin-Bona-Mahony equations or the modified Korteweg-de Vries equations}

\subsubsection{The nonlinear dispersive Benjamin-Bona-Mahony equation or the modified Korteweg-de Vries equation. \;\; Version 1 [26, 71]}


The nonlinear dispersive modified Benjamin-Bona-Mahony equation was derived to describe an approximation for surface long waves in dispersive nonlinear materials, with phonon properties in the anharmonic crystal lattice, acoustic-gravity waves in compressible fluids, and hydro-magnetic waves in cold plasma.

The generalized  modified Benjamin-Bona-Mahony equation has the form [67-70, 87, 88].
\begin{equation}\label{mbbm}\nonumber\\
u_{ t}+C u_{z}+\beta u_{zzz} +a u^{n} u_{z}=0.
\end{equation}
When $n=2$ this equation called the  nonlinear dispersive  modified Benjamin-Bona-Mahony equation
\begin{equation}\label{mbbm}
u_{ t}+C u_{z}+\beta u_{zzz} +a u^{2} u_{z}=0,
\end{equation}
where $C$, $\beta$ and $a$ are arbitrary constants.

Eq.(27) have been considered as an improvement of the modified Korteweg-de Vries equation  which is also well known in different field of research and applications [68-70, 87, 88].

\subsubsection{The nonlinear dispersive Benjamin-Bona-Mahony equation or the modified Korteweg-de Vries equation. \;\;\;Version 2 [89, 90]}

Sometimes, another form of the  modified Benjamin-Bona-Mahony equation when the term $u_{zzz}$ is replaced by the term $u_{zzt}$  is also considered [91-95]
\begin{equation}\label{bbm}
u_{ t}+C u_{z}+\beta u_{zzt} +a u^{2} u_{z}=0.
\end{equation}

The formation of nonlinear waves in various physical systems (anharmonic crystals, hydrodynamics, plasma, etc.) that is described by the  modified Benjamin-Bona-Mahony equation (28)

\subsection{Nonresonanse vector breather in Kerr and dispersive media [96]}

We consider the optical nonresonance two-component vector breather in a dispersive medium with a third-order Kerr
susceptibility for linearly polarized waves with the $x$  component
strength of the electric field $\vec{u}=(u,,0,0)$  propagating along the positive $\emph{z}$-axis,
The nonlinear wave equation has the following form [97, 98]:
\begin{equation}\label{nonre}
 c^2 u_{zz} -D_{tt} -4\pi (P_{nr})_{tt} =0,
\end{equation}

where
$D=\int \kappa (z_1,t_1) u(z-z_1,t-t_1)dz_1 dt_1$ is linear part of $x$ component of the electric induction vector.
\begin{equation}\label{nonker}
P_{nr}=\int\rho_{xxxx}({z}_1,{z}_2,{z}_3,t_1,t_2,t_3)u(z-z_1,t-t_1)\times $$$$
u(z-z_1-z_2,t-t_1-t_2)u(z-z_1-z_2-z_3,t-t_1-t_2-t_3) dz_1 dz_2
dz_3 dt_1 dt_2 dt_3
\end{equation}
is an $x$-component of the nonresonant nonlinear polarization of the Kerr-type medium and $\rho_{xxxx}$  is a
susceptibility tensor component, $c$ is  the light velocity in vacuum.

\subsection{The nonlinear equation for two-photon SIT [99-102]}

The Maxwell nonlinear wave equation for two-photon SIT has the form
\begin{equation}\label{twoph}
-c^2 u_{tt} + u_{zz} = 4\pi   P^{(2p)}_{tt},
\end{equation}
where $P^{(2p)}$ is the  the resonance polarization of two-level optical impurity
atoms or SQDs under the condition of the two-photon resonance transitions.

 Eq.(31) describes nonlinear coherent interaction between an optical pulse and material in which resonance optical impurity atoms or SQDs have been embedded. In particular, under the condition of the two-photon SIT.

\vskip+0.5cm

\subsection{Hybrid equation [103, 104]}

Depending  on  the  numerical  values  of light and medium  parameters there may occur  physical  situations in which  both  resonant and nonresonant mechanisms  of the formation  of nonlinear  waves act   simultaneously. In that case, the  hybrid  mechanism  of excitation of a  nonlinear  solitary  wave becomes active and a hybrid  nonlinear  pulse  can be formed.

We consider the propagation of optical hybrid two-component vector breather in a dispersive medium with a third-order Kerr
susceptibility containing a small concentration of optically active impurity atoms or SQDs.

The wave equation for the $\emph{x}$ -component of the strength of the electric field $\vec{u}(u,0,0)$  is written in the form [97, 98]

\begin{equation}\label{eq}
 c^2 u_{zz} -D_{tt} -4\pi (P_{nr}+P_{r})_{tt} =0.
\end{equation}
The nonlinear polarization of the medium contains the nonresonant $P_{nr}$ and resonant $P_{r}$ parts. They are determined from the equations (30) and (B2) respectively.

\vskip+0.5cm

\subsection{Acoustic SIT equation [28]}

The equation of the theory of elasticity for the transverse-polarized  acoustic pulse under the condition of acoustic SIT has the form
\begin{equation}\label{asit}
u_{tt}=c^{2}_{t}u_{zz}+
 \frac{\beta_{0} H_{0} F_{xzxz} n_{p}}{2 \rho} (s_{x})_{zz},
\end{equation}
where $u=\varepsilon_{xz}$ is component of the deformation tensor [105], $\rho$ is the density of the medium, $c_{t}$ is the velocity of the  transverse-polarized linear acoustic wave  in medium, $\beta_{0}$ is the Bohr magneton, $F_{xzxz}$ is component of the spin-phonon coupling tensor, $H_{0}$ is the external
constant magnetic field. $n_p$ and ${s}_{x}$ are the concentration and the average value of the $x$ component of the electron spin-operator of the paramagnetic impurity atoms, respectively.

\subsection{General equation of the fourth-order [101, 102]}

In view of the fact that the same phenomenon, namely the formation of a vector breather consisting of two breathers oscillating with the SDFW, can be observed in completely different areas of physics, for quantities of completely different physical nature, which are described by different nonlinear equations, it follows that the existence of such vector breathers is a general fundamental characteristic of matter. Consequently, a natural question arises: is it possible to describe such vector breathers in completely different areas of physics using one general equation?

In this section, precisely such an equation containing the second and fourth-order derivatives in the spatial coordinates and time is proposed that makes it possible to describe vector breathers from various fields of physics in a unified way.

The general equation containing the second and fourth-order derivatives in the spatial coordinates and time which describe pulses can take on the following form:
\begin{equation}\label{geneq}
\alpha u_{tt} + \beta u_{zz} + \gamma u_{zt} + \delta u_{zzzz}+ \mu u_{zzzt} + \nu u_{zztt}+ \rho u_{zttt}+ \vartheta u_{tttt} =F_{non},
\end{equation}
where the nonlinear term of Eq.(34) is given by
\begin{equation}\label{gennon}\nonumber
F_{non}=- G   (u^3)_{zz} -  A (u_{t})^{2} u_{ zz}  - \sigma (u_{z})^{2} u_{ tt} + B u_{t} u_{z} u_{zt}-\alpha_{0}^{2} \sin u  +F  P^{(2P)}_{tt},
\end{equation}
$\alpha,\;\beta,\;\gamma,\;\delta,\;\mu,\;\nu,\;\rho$, $\vartheta$, $G$, $A$, $\sigma$, $B$, $\alpha_{0}^{2}$ and $F$ are arbitrary constants.

The nonlinear wave equation (34) is a general equation which depending from the coefficients has several special cases.

1. When  the coefficients in Eq. \eqref{geneq} satisfied the conditions
\begin{equation}\label{c1}\nonumber
\alpha=1,\;\;\beta=-C,\;\; \gamma=\delta=\mu =\nu =\rho =\vartheta=G=A=\sigma=B=F=0,
\end{equation}
this equation is transformed into the SGE Eq.(13).

2. In the case when
\begin{equation}\label{cp}\nonumber
\alpha=1,\;\;\beta=-C,\;\; \gamma=\delta=\mu =\nu =\rho =\vartheta=G=F=\alpha_{0}^{2}=0,
\end{equation}
Eq.(34) is reduced to the Born-Infeld equation (15).

3. In the case when
\begin{equation}\label{cp}\nonumber
\rho =\vartheta=A=\sigma=B=F=\alpha_{0}^{2}=0,
\end{equation}
Eq.(34) is transformed to the nonlinear cubic Boussinesq-type equation (23).

4. When  the coefficients satisfied the conditions
\begin{equation}\label{c1}\nonumber
\alpha=1,\;\;\beta=-\frac{1}{c^2},\;\;F=-\frac{4\pi}{c^2},\;\; \gamma=\delta=\mu =\nu =\rho =\vartheta=G=A=\sigma=B=0,
\end{equation}

Eq.(34) is reduced to the Maxwell nonlinear wave equation for two-photon SIT Eq.(31).

\vskip+0.5cm

\subsection{Vector $0\pi$ pulse and some other nonlinear equations}

These are links to two-component vector breathers - vector $0\pi$ pulses  in some other physical systems and for for some other physical phenomena [18, 25, 30, 37, 48, 99, 106-139]. They include, in particular, the nonlinear equations for surface plasmon polaritons, for optical waveguide modes, for the extraordinary optical waves in anisotropic uniaxial media, for acoustic waves in anisotropic media, for nonlinear surface acoustic waves, such as Rayleigh waves, Love waves, and generalized Love waves. Nonlinear equations in quantum dots, matamaterials, graphene, etc.

Eqs.(10) - (34) and also equations mentioned in Refs.[18, 25, 30, 37, 48, 99, 106-139] have a solution in the form of a two-component vector breather oscillating with the SDFW - vector $0\pi$ pulse:
\begin{equation}\label{VB2}
u(z,t)=
\frac{2}{{b} T}sech(\frac{t-\frac{z}{V_{0}}}{T})\{  K_{+} \cos[(k+\kappa_{+})z -(\om + w_{+}) t] +K_{-}\cos[(k-\kappa_{-})z -(\om -w_{-})t]\}.
\end{equation}
The relations between the parameters are presented in Appendix C.

The GPRM is applied to various physical quantities. Therefore, the explicit analytical form of the vector $0\pi$ pulse may differ slightly, as, for example, in Eqs.(9) and (35). This is because in one case, the GPRM is applied to the function characterizing the physical process, while in the other case, the GPRM is applied to the integral (or derivative) of the quantity characterizing the physical process.
For example, for the  modified Benjamin-Bona-Mahony equation, the Boussinesq equation, and other equations presented in this Section the GPRM is applied to the functions characterizing the corresponding physical process, while in the case of the SIT, it is applied to the integral of the quantity characterizing the physical process Eq.(C1).
The difference between these expressions for the vector $0\pi$ pulses lies in the renormalization of the amplitude and the phase shift. The corresponding graphical expressions for vector $0\pi$ pulse in the both cases are identical (Fig.1).

All nonlinear equations listed in this section are solved using the GPRM Eq.(3) and have the solution in the form of a vector  $0\pi$ pulse Eq.(35),  with the same wave profile (Fig.1), differing from each other in the values of the parameters $p_{\pm},$$\; q_{\pm},\;$$ r_{\pm}$ and $v_{\pm }$, as well as the dispersion relation between the frequency $\omega$ and wave number $k$ of the carrier wave and the relation between the oscillating parameters $\Omega_{\pm}$ and $Q_{\pm}$. The explicit analytical expressions of these parameters for the each of mentioned nonlinear equations are given in the relevant references.

\section{Conclusions}

In this review, using the recently developed  approach  GPRM, we pursued two main goals: first, to show that the more general theory of the SIT significantly differs from the original theory of McCall and Hahn. Namely,  the fundamental pulse of the SIT is a vector  $0\pi$ pulse but not of a scalar  $0\pi$ pulse of McCall and Hahn, and second, to show that the vector  $0\pi$ pulse is a universal nonlinear wave that occurs in almost all areas of physics and is a solution to a number of nonlinear equations.

In the SIT wave equation the first derivative of the strength of the electric field of the pulse with respect to the spatial coordinate and time,  can describe the formation of the scalar soliton and the scalar breather. The corresponding second derivatives terms in the nonlinear wave equation (B4)  are smaller than the first derivatives and describe the interaction of scalar breathers and the formation of their bound state - the two-component vector breather oscillating with the SDFW (vector $0\pi$ pulse).

When solving nonlinear differential equations using reduction methods, the original equation is transformed into another equation with respect to auxiliary functions, the solution to which is well known. Using the standard version of the PRM [14, 15], a wide class of nonlinear differential equations is reduced to a scalar NSE with respect to auxiliary function. A similar situation occurs for the GPRM Eq.(3), in which a fairly wide class of nonlinear differential equations is transformed into a VNSEs with respect to auxiliary functions, the solution to which also is well known.

Considering that the same phenomenon, namely the formation of a vector breather consisting of two breathers oscillating with the SDFW, can be observed in completely different areas of physics, for quantities of completely different physical nature, which are described by different nonlinear equations, it follows that the existence of such vector breather (vector $0\pi$ pulse) is a general fundamental characteristic of matter along with scalar soliton and scalar breather.

\vskip+0.2cm

\section{Abbreviation}

SIT - self-induced transparency

IST - inverse scattering transform

PRM - perturbative reduction method

GPRM - generalized perturbative reduction method

SQDs - semiconductor quantum dots

NSE - nonlinear Schr\"odinger equation

VNSEs - vector nonlinear Schr\"odinger equations

SDFW - sum and difference of the frequencies and wave numbers

SVEA - slowly varying envelope approximation

SGE -  sine-Gordon equation

\section{Appendices}

\appendix


\section{Scalar solitons and scalar breathers }

\label{MH SIT}

This appendix presents a list of various research directions of resonant, non-resonant and hybrid single-component (scalar) solitons and single-component (scalar) breathers:
\vskip+0.2cm

Hybrid nonlinear waves [118, 140, 141];

Optical scalar breathers [20-24, 32, 107, 108, 113, 142, 145-147, 152-160, 190, 201, 220];

Acoustic solitons and scalar breathers [109-111, 170, 178, 194, 198, 200, 202-207, 242-245];

SIT in SQDs [35, 36, 48, 147, 201, 208-212, 224];

Influence of relaxations on nonlinear waves [19, 32, 43, 46, 111, 170, 174, 180, 183, 187, 199, 207, 213-215];
 
 Nonlinear surface plasmon polaritons and waveguide modes [18. 108, 189, 201, 208, 214, 219, 220, 238-240];

Nonlinear waves under the condition of dissipation [221];

Nonlinear waves in graphene and phosphorene [35, 37, 108, 143, 148, 201, 222, 223];

SIT of excitons in anisotropic media [161];

Acoustic SIT in anisotropic media [169];

SIT for EPR region of spectra [167, 195, 197];

SIT in magneto-ordered media [172, 179, 188];

Acoustic SIT under the condition of the electron-nuclear excitations [169, 181];

Polarization of optical waves in resonance-absorbtion media [165];

Polarization of acoustic soliton in paramagnetic crystal [166];

Polarization of magneto-acoustic waves in ferromagnets [185];

SIT in anisotropic media [21, 22, 24, 25, 33, 114, 144, 164, 174, 182];

Nonlinear waves in metamaterials [214, 220, 226, 227, 232, 236, 237];

Nonlinear wave in dispersive media [23, 24, 106, 108, 112, 163];

SIT under the condition of the two-photon resonance transitions [22, 43, 99, 162, 163, 241];
 
SIT under conditions of excited forbidden optical transitions [186]; 

Nonlinear surface acoustic waves [110, 168, 171, 173, 175-177, 184, 191, 193, 196, 202, 204, 205, 216, 217, 225, 228-231, 233-235];

Acoustic SIT under the condition of the two-phonon resonance transitions [175, 177 189, 192, 218, 228];

Acoustic nonlinear wave in anisotropic media [109, 192];

Nonresonance optical breathers [23];

Optical and acoustic nonlinear waves experiments [19, 20, 32-34, 38-43, 149-151, 242-245].

\vskip+0.2cm

\section{ The Maxwell-Bloch equations}

\label{BME}

We will consider an optical pulse that is linearly polarized along the $x$ axis and that propagates along the $z$ axis in a medium that contains a small concentration $n_{0}$ of impurity optically active two-level atoms.

The Maxwell wave equation for the x component of the electric field strength $E$ of the wave has the form
\begin{equation}\label{maxgen}
c^2{\partial}_{zz} E- \ve_{0}{\partial}_{tt} E = 4\pi {\partial}_{tt} P,
\end{equation}
where  $c$ is speed of light in vacuum, $\ve_{0}$ is the dielectric permittivity,  $P=n_{0} {\mu}_{12} s_{x}$ is the polarization of two-level
impurity atoms, and
$s_{i}=Tr<\Psi|\hat{\sigma}_{i}|\Psi>$,  are average values of the Pauli operator $\hat{\sigma}_{i}$ for the state $|\Psi >,\;\;i=x,y,z.$
They satisfy the optical Bloch equations [246]

\begin{equation}\label{bloch}
{\partial}_{t} s_{x}=- \omega_{0} s_{y},
$$$$
{\partial}_{t} s_{y}= \omega_{0} s_{x}+ \kappa_{0} E s_{z},
$$$$
{\partial}_{t} s_{z}=- \kappa_{0} E s_{y},
\end{equation}
where  $\omega_{0}$ is the resonant excitation frequency of impurity two-level atoms; $\kappa_{0}=\frac{2{\mu}_{12}}{\hbar}$ ,  ${\mu}_{12}$  is the electric dipole moment for the corresponding transition  of an impurity atom, assumed to be real. $\hbar$ is the Planck constant. We
assume that $T<<T_{1,2}$, in Eq.(B2) and then neglect the relaxation effects of the impurities.

When the pulse width is longer than a few periods of the carrier wave $ T\gg 1/\omega $, the study of SIT can be done with the SVEA for the electric field strength $E$ of the wave Eq.(1) (where we substitute $E$   instead of $u$) and analogously the polarization of two-level impurity atoms $P$ we can represent as
\begin{equation}\label{pl}
P=\sum_{l=\pm1}p_{l}Z_l,
\end{equation}
where $p_{l}$ is the slowly varying complex amplitudes of the polarization, which satisfied the inequalities
$$
|\partial_{t} p_{l}|<<\omega
|p_{l}|,\;\;\;|\partial_{z} p_{l}|<<k|p_{l}|.
$$

Upon taking into account inhomogeneous broadening of the spectral line, the polarization of the two-level system is equal to
\begin{equation}\nonumber
P=i \mu_{12} n_{0} \int g(\Delta)( \rho^{+} Z_{-1}- \rho^{-}Z_{1})d \Delta,
\end{equation}
where $g(\Delta)$ is the inhomogeneous broadening function, $\Delta=\om_{0}-\om\;$.
$\rho^{+}= v^{*}_{1}v_{2},\;\;\rho^{-}= v_{1}v^{*}_{2},$
the functions $v_{1}$ and $v_{2}$ and their complex conjugate functions are solutions of the Zakharov-Shabat equations [3, 5, 44].

Substituting Eqs.(1) and (B3) into Eqs.(B1) and (B2), we obtain the equations for the slowly varying complex envelope of the electric field $\hat{E}_{l}$ and the undamped Maxwell-Bloch equations for the slowly varying complex envelope of the polarization $\rho^{\pm}$ in the form
\begin{equation}\label{maxslow}
\sum_{l=\pm1}Z_l (-2il\omega {\partial}_{t} \hat{E}_{l}
- 2ilk v^{2} {\partial}_{z} \hat{E}_{l} +{\partial}_{tt} \hat{E}_{l} -v^{2}
{\partial}_{zz} \hat{E}_{l}- \frac{4\pi {\omega}^{2}}{\ve_{0}}  p_{l})=0,
\end{equation}

\begin{equation}\label{blochslow}
{\partial}_{t} \rho^{+}= i\Delta  \rho^{+} -  r s_{z},
$$$$
{\partial}_{t} \rho^{-}= -i\Delta \rho^{-} +  q s_{z},
$$$$
{\partial}_{t} s_{z}=2(r \rho^{-}-q \rho^{+}).
\end{equation}
where

\begin{equation}\label{hh}\nonumber
v^{2}=\frac{c^2}{\ve_{0}},\;\;\;
r=-\frac{{\mu}_{12}}{\hbar }\hat{E_{1}},\;\;\;q=\frac{{\mu}_{12}}{\hbar } \hat{E^{*}_{1}},\;\;\; r=-q^*.
\end{equation}

Usually, the simplest way, at the solution of the system of equations (B4) and (B5), when the wave equation  contains only the first-order derivatives in the spatial coordinate $z$ and time $t$, while the corresponding second-order derivatives have been neglected. In such case, Eq.(B4) is reduced to the following nonlinear wave equation for the envelope function:
\begin{equation}\label{rmax}
{\partial}_{z} r +\frac{\eta }{c}  {\partial}_{t} r =- \frac{2\pi n_{0}{\mu}^{2}_{12}{\omega} }{c  \eta \hbar} \int g(\Delta)\rho^{+}d  \Delta,
\end{equation}
where  $\eta$ is a reflective index $(\eta^{2}=\ve_0)$.

In this special case the undamped Bloch-Maxwell equations (B5) and (B6), after ignoring the phase modulation and the second-order derivatives with respect to spatial coordinates and time, is transformed to the SGE Eq.(13), when $u={\Psi}$,
where the area of the pulse envelope
\begin{equation}\label{area}
{\Psi}(z,t)=\kappa_{0} \int_{-\infty}^{t}\hat{E}(z,t')dt'.
\end{equation}

\vskip+0.5cm

\section{The vector nonlinear Schr\"odinger equations}

\label{VNSE}

Upon phase modulation, the function $\Psi_{l}$  is complex function. When the condition of $|\Psi_{l}|<<1$ is fulfilled, this function can be represented in the form
\begin{equation}\label{Psi}
\Psi_{l}(z,t)=\kappa_{0} \int_{-\infty}^{t}\hat{E}_{l}(z,t')dt'=\sum_{\alpha=1} \ve^\alpha
{{\Psi}_{l}}^{(\alpha)}(z,t).
\end{equation}

Substituting of Eqs.(3) (when $\hat{u}_l=\Psi_{l}$)  and (C1) into Eq.(B4) we obtain the equation
\begin{equation}\label{wjhH}
\sum_{l=\pm1}\sum_{\alpha=1}\sum_{n=-\infty}^{+\infty}
  \ve^\alpha  Z_{l} Y_{l,n}\{\tilde{W}_{l,n} +\ve J_{l,n} {\partial }_{\zeta}
 +\ve^2 h_{l,n} {\partial }_{\tau}+
{\ve}^{2}i H_{l,n} {\partial}_{\zeta \zeta}\}f_{l,n}^ {(\alpha)}$$$$=-\ve^{3}i \frac{\beta^{2}}{2}\sum_{l=\pm 1}l Z_l
 \int {\partial}_{t}{{\Psi}_{l}}^{(1)}    {{\Psi}_{-l}}^{(1)} {{\Psi}_{l}}^{(1)}dt'
 +O(\ve^4),
\end{equation}
where
\begin{equation}\label{wjhH1}
{W}_{l,n}=in \Omega (A_{l}n {\Omega} - B_{l} n Q  +
{\Omega}^{2}-v^{2}  Q^{2}  -\frac{l}{n }\frac{\beta^{2}}{ \Omega}),
$$
$$
J_{l,n}= n Q[2A_{l} \Omega
v_g -B_{l}(Q v_g +\Omega) + 3n {\Omega}^{2}  v_g -v^{2}n Q(Q v_g + 2
\Omega)],
$$
$$
h_{l,n}=-2 n
A_{l}\Omega + B_{l}nQ - 3{\Omega}^{2} +v^{2}   Q^{2},
$$
$$
H_{l,n}=  Q^{2} [-A_{l}   v_g^{2} + B_{l}   v_g
-3n\Omega  {v_g}^{2}+ v^{2} n ( 2Q v_g +\Omega)],
$$$$
A_{l}=2l\omega,\;\;\;\;\;\;\;\;\;\;\;\;\;\;\;\;B_{l}=2lkv^{2},\;\;\;\;\;\;\;\;\;\;\;\;\;\beta^{2}=\frac{4 \pi {\omega}^{2} n_{0} {{\mu}_{12}}^{2}
 }{\hbar \ve_{0}}  \int \frac{g(\Delta )d \Delta}{1+\Delta^{2}T^{2}},
$$
$$
\om^{2}=v^{2}k^2 .
\end{equation}

Here, for the sake of simplicity, we omit $l$ and $n$ indexes for the quantities  $\Omega_{l,n}$, $Q_{l,n}$, ${{v}_{g;}}_{l,n}$ and $\zeta_{l,n}$.

Following the standard procedure characterized for any perturbative expansions while equating to each other the terms of the same order to $\ve$ from the equations (C2) and (C3)
we obtain the chain of the equations. As a result, we determine the connection between the oscillating parameters Eq.(5)
and the VNSEs Eq.(6).

Nonlinear equations (6) are the base equations of the theory of nonlinear two-component waves to describe the dynamics of the envelopes of $U_{+}$ and $U_{-}$ of a one-dimension wave
packet.
There are many different methods for solving VNSEs. Let's choose one of them.

We will seek the steady-state solutions of Eq.(6) for the functions $U_{\pm}$ in the following form
\begin{equation}\label{uu2}
U_{\pm}=K_{\pm}\; S( \xi )e^{i\varphi_{\pm}},
\end{equation}
where $\varphi_{\pm}=k_{\pm} z- \omega_{\pm} t$ are the phase functions for each components, $K_{\pm},\;k_{\pm}$ and $\omega_{\pm}$ are constants.
The standard way to ensure the stationary character of the envelope function $S( \xi )$ entails depending on spatial coordinate and time only through the variable $ \xi = t- \frac{z}{V_{0}}$, where $V_{0}$ is the nonlinear two-component pulse velocity. We consider that valid following inequalities
$ k_{\pm}<<Q_{\pm },\;\;\;\;\omega_{\pm}<<{\Omega}_{\pm }.$

Substituting Eq.(C4) into Eq.(6), we obtain the explicit form of the envelope function
\begin{equation}\label{rt16}
S(\xi)=\frac{1}{{b} T}sech(\frac{t- \frac{z}{V_{0}}}{T}),
\end{equation}

Substituting Eqs.(C4) and (C5) into Eqs.(3) and (1), we obtain Eq.(9), where the connections between different parameters of the nonlinear wave
are given by
\begin{equation}\label{par}
K_{+}^{2}=\frac{p_{+}q_{-}-2 p_{-} q_{+}}{p_{-}q_{+}-2p_{+} q_{-}}K_{-}^{2},
\;\;\;\;\;\;\;\;\;\;\;\;\;
\omega_{+}=\frac{p_{+}}{p_{-}}\omega_{-}+\frac{V^{2}_{0}(p_{-}^{2}-p_{+}^{2})+v_{-}^{2}p_{+}^{2}-v_{+}^{2}p_{-}^{2}
}{4p_{+}p_{-}^{2}},
$$$$
k_{\pm}=\frac{V_{0}-v_{\pm}}{2p_{\pm}}, \;\;\;\;\;\;\;\;\;\;
T^{-2}=V_{0}^{2}\frac{v_{+}k_{+}+k_{+}^{2}p_{+}-\omega_{+}}{p_{+}},\;\;\;\;\;\;\;\;\;\;\;
{b}^{2}=\frac{V_{0}^{2} q_{+}}{2p_{+}}(K_{+}^{2}+2 K_{-}^{2}) .
\end{equation}

\vskip+0.5cm

\end{document}